\documentclass[fleqn,twoside]{article}
\usepackage{amsmath,espcrc2}

\usepackage{graphicx,epsfig}
\usepackage[figuresright]{rotating}

\title{Complex-band structure:
a method to determine the off-resonant electron transport in oligomers}

\author{Giorgos Fagas
\address[RU]
{Institut f\"{u}r Theoretische Physik, Universit\"{a}t Regensburg\\
D-93040 Regensburg, Germany}
        \thanks{Corresponding author:\newline
                giorgos.fagas@physik.uni-regensburg.de
               },
Agapi Kambili
\addressmark[RU]
, and
Marcus Elstner
\address[PU]
{Theoretische Physik, Universit\"{a}t
Paderborn, D-33098 Paderborn, Germany , and \\
Abteilung f{{\"u}}r Molekulare Biophysik, Deutsches
Krebsforschungscenter, D-69120 Heidelberg, Germany}
}
       
\begin{document}

\begin{abstract}
\vspace{1pc}
We validate that off-resonant electron transport across {\it ultra-short}
oligomer molecular junctions is characterised by a conductance
which decays exponentially with length, and we
discuss a method to determine the damping factor via the energy
spectrum of a periodic structure as a function of complex wavevector.
An exact mapping to the complex wavevector is demonstrated
by first-principle-based calculations of:
a) the conductance of molecular junctions of phenyl-ethynylene wires
covalently bonded to graphitic ribbons as a function of the bridge length,
and b) the complex-band structure of poly-phenyl-ethynylene.

\end{abstract}

\maketitle

\section{Introduction}
Since the pioneering theoretical proposal of a molecular rectifier by
Aviram and Ratner \cite{CPL74AR},
molecular electronics has received considerable attention
\cite{Sci03NR}.
A first key experiment \cite{JAP71MK} investigated the
predictions of tunnel theory. It established the exponential
decrease of the conductance in Langmuir-Blodgett
[CH$_3$(CH$_2$)$_{N-2}$COO]$_2$Cd films between Al electrodes versus
their thickness for chain lengths $N$=18-22.
Most recently, in an effort to illuminate the mechanism of
electron transport in relatively short mono-molecular junctions,
such length dependence has been reported in experiments with scanning
probe microscopy techniques
\cite{APL01SHI,JPCB02WHR,JPCB02IMA,JPCB02CPZ},
where a small number of molecules contributes to the conductance.
Structure dependent factors have been studied by looking at a variety
of oligomers, and by systematically varying the
contact to the electrodes via the anchor groups or the tip load.
In some cases, measured molecules had chain lengths ranging from one
up to four monomer units \cite{APL01SHI,JPCB02WHR,JPCB02IMA}.
All results point to through-bond off-resonant tunnelling transport.

In metal-molecule-metal junctions (MMMs),
the molecular electronic states hybridise
with the metal wavefunctions giving a finite width in the energy space.
If no large charge transfer occurs from/to the molecule, the Fermi energy
$E_F$ lies within the (broadened and shifted) highest-occupied- and
lowest-unoccupied- molecular-orbital (HOMO-LUMO) energies. Such an
alignment poses a potential barrier to traversing electrons accounting
for tunnelling.
In this regime, there is an analogy of MMMs to metal-insulator
(or metal-semiconductor) junctions, and of the molecular-orbital
energy broadening to the metal-induced gap states (MIGS)
\cite{PRB65H,PRB601LA,PRL00LBS,PRB02TS}.

Such states are induced by matching to the metal side, appear at energies
within the band-gap of the insulator (semiconductor), and decay exponentially
away from the junction in the non-metal \cite{PRB65H}. MIGS are responsible
for the overall finite density of states (DOS) within the HOMO-LUMO gap $E_g$
for short molecular wires because of contributions coming from both
electrodes \cite{PRB601LA}. For longer wires, one expects the MIGS to be
mainly located near the interfaces on either side of the molecule.
By looking at the
positional dependence of the local DOS, such behaviour has been observed for
Si nanowires between Al electrodes \cite{PRL00LBS} and octaneditiols
between Au (111) surfaces \cite{PRB02TS}. In the latter case, a one-to-one
correspondence between the local DOS decay parameter and the complex wavevector
of wavefunctions in the forbidden energy domain of the corresponding
one-dimensional crystal, was found.

Such a property has been conjectured to also hold for the damping factor $\beta$ of
the conductance of oligomer molecular junctions \cite{PRB02TS,SSC98OKM,CP02JM},
deep in the tunnelling regime, where $\beta N\gg1$.
The physical meaning of $\beta$ arises naturally, and the dispersion relation with
complex wavevector may be used to define an effective mass of tunnelling
electrons \cite{PRB02TS,CP02JM}. However, since most studies assume
that $\beta N\gg1$, the range of validity of such tunnelling characteristics is unclear.
In what follows, we elaborate on this and demonstrate that the conductance
decays almost always with length in an exponential fashion.
We focus on the damping factor of electron transport across
oligomers which is the relevant quantifying quantity, readily deducible
experimentally. In particular, we establish its
connection to the band structure of the corresponding polymer when extended
in the complex wavevector plane. Most importantly,
this mapping holds even for very short bridges consisting of a handful of monomer units,
as in experiments, and for surprisingly small values of the damping factor,
providing an indispensable tool for the understanding and prediction of the
tunnelling phenomena involved.

\section{Electron transport framework}
\label{s2}
The low-bias and low-temperature electronic
response of the MMMs is determined by the elastic
scattering of independent electrons with energy $E$ for
molecular wires in which electron resident times are
small compared to those of molecular vibrations \cite{Sci03NR}.
Electrons see a frozen-nuclei configuration and,
in a first approximation, the main contribution derives
from equilibrium conditions. Polaronic effects may also be
disregarded for molecular bridges of the size we consider \cite{PRB01NSF},
whereas, the response is voltage independent \cite{CP01MR}.
The Landauer conductance $\mathrm g$ is
\begin{equation}
{\mathrm g}(E)=\frac{2e^2}{h} \; T(E),
\label{1}
\end{equation}
where the transmission function $T$ denotes the total transition
probability from incoming to outgoing states at mutually exclusive
electrodes. The
temperature constrain is relaxed by comparing the molecular
energy scales ($\sim$ eV) to room temperature ($\approx$ 25meV).

\vspace{-0.5cm}
\begin{figure}[h]
\centerline{\epsfig{figure=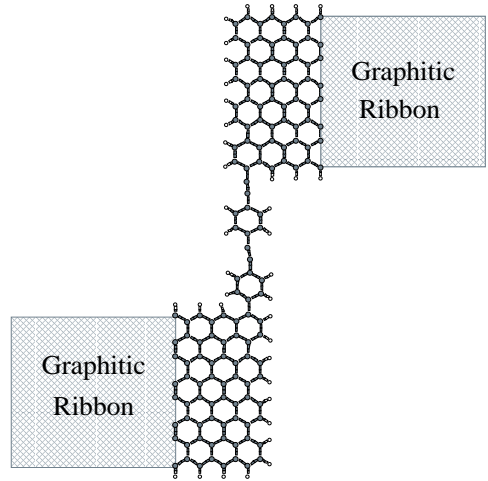,width=15pc,height=15pc}}
\vspace{-0.75cm}
\caption{Structure of diphenyl-ethynylene bonded to two
graphitic electrodes. Open circles denote hydrogen atoms.
See Sec.\ref{s4} for more details.}
\label{f1}
\end{figure}
\vspace{-0.5cm}

For energies far-off the molecular bridge resonant-spectrum
and relatively long wires,
the transmission function has been studied in both model
systems and realistic molecular junctions
\cite{SSC98OKM,JCP94MKR,PRB98MJ,JCP00HRH,EPL96JV}.
It shows the characteristic exponential length dependence of a tunnelling
process
\begin{equation}
T(E) = T_o(E)  \; {\mathnormal e^{- \beta(E) (N-1) }},
\label{2}
\end{equation}
where $N$ represents the number of monomers that constitute the
molecular wire. In Sec.\ref{s5},
we show that Eq.\ref{2} is of more general validity.

The pre-exponential factor relates to the molecule/electrode
interface; it defines the contact conductance
${\mathrm g}_o$ in units of $(2e^2/h)$. In the long oligomer limit,
the damping factor $\beta$ is a property of the bridge, related to
its electronic structure. Based on WKB approximation
a simple expression reads
\begin{equation}
\beta(E) = 2\sqrt {\frac{2m}{\hbar^2}{E}_{\rm g\prime}},
\label{3}
\end{equation}
where $E_{\rm g\prime}$ equals $E_{\rm g}$ or $|E-E_{\rm HOMO(LUMO)}|$.
However, this overestimates $\beta$ \cite{PRB02TS,CP02JM} and calls for an
elaborated approach as discussed later.

To investigate the conductance properties of
molecular junctions as that
of Fig.\ref{f1},
we calculate the transmission function with Green function
methods of quantum transport \cite{Datta} via
\begin{equation}
T(E)=Tr\left[{\mathbf \Gamma_L(E)}{\mathbf G^r(E)}
{\mathbf \Gamma_R(E)}{\mathbf G^a(E)}\right].   
\label{4}
\end{equation}
${\mathbf G}^{r(a)}=(E \mathbf S-\mathbf H- {\mathbf \Sigma)^{-1}}$ is
the retarded (advanced) molecular Green function dressed by a
self-energy interaction ${\mathbf \Sigma} = {\mathbf \Sigma_L} + {\mathbf \Sigma_R} $.
This follows the L{{\"o}}wdin projection technique
and takes into account the coupling to the 'left' (L) and 'right' (R)
electrodes. Twice the imaginary part of ${\mathbf \Sigma_{L(R)}}$ defines
the spectral width ${\mathbf \Gamma}_{L(R)}$. Since we are interested in
the low-voltage regime we use the equilibrium Green function.
We have assumed a non-orthogonal basis which
introduces the overlap matrix $\mathbf S$, and an algebraic
procedure with finite dimensional matrices.

To treat the electronic structure of the oligomer and the electrodes,
we employ a linear combination of atomic orbitals approach
parameterised by Density Functional Theory (DFT-TB)
\cite{PRB95PFK} in the local density approximation (LDA).
In DFT-TB, the
single-particle electronic Kohn-Sham eigenstates $\psi_i$ of the system
are expanded in a non-orthogonal basis set $\varphi_\mu$ for
valence electrons located at the ionic positions ${\mathbf R_\mu}$, namely
\begin{equation}
\psi_i({\mathbf r})=\sum\limits_\mu  c^i_\mu \varphi_\mu({\mathbf r-R_\mu }).
\label{5}
\end{equation}
The basis set is computed in terms of Slater-type orbitals
in a modified atomic potential that slightly compresses
the electron density to take into account that atoms are brought
together in a condensed state. With the {\it Ansatz} of Eq.\ref{5}
the Kohn-Sham equations for $\psi_i$ are transformed to a set
of algebraic equations
\begin{equation}
\sum\limits_\nu (H_{\mu \nu }-S_{\mu \nu } E_i)c^i_\nu =0.
\label{6}
\end{equation}
The tight-binding (TB) approximation assumes superimposing
atomic densities when calculating potential contributions
\cite{PRB98EPJ}. $H_{\mu \nu }$ include only two-centre,
distance-dependent, Hamiltonian matrix elements. Eq.\ref{6} is
of the extended H\"{u}ckel type but all
elements are calculated numerically without
empirical parameters. Moreover, the a priori
parameterisation of $\mathbf H$ and $\mathbf S$ provides
an efficient scheme for calculating the Green functions.

\section{Complex-band structure}
\label{s3}
Assuming that interfacial effects are not dominant, any
evanescent states in an underlying periodic system
are understood as bulk Bloch states with a $k$ vector with an imaginary
component, i.e., $k = q - \imath \kappa$. These states
are quantified by the complex-band structure, the extension
of the electronic band structure to include complex Bloch
vectors. In band structure calculations, real $k$ in the
first Brillouin zone are given as input to solve the
energy eigenvalue problem.
For the complex-band structure, the inverse problem is
defined. All $k$ vectors associated with a real energy $E$ are
sought, which are in general complex. The analytic
properties of the energy spectrum of crystals as function of a
complex variable have been initially studied by Kohn \cite{PR59K}.

In any case, Eq.\ref{6} in a periodic potential is solved,
which within our TB approximation reads
\begin{equation}
\begin{split}
&(\mathbf H_{m,m}+\mathbf H_{m,m+1}e^{ika}+\mathbf H_{m-1,m}e^{-ika})\phi_k^E= \\
&E_k(\mathbf S_{m,m}+ \mathbf S_{m,m+1}e^{ika}+\mathbf S_{m-1,m}e^{-ika})\phi_k^E.
\label{7}
\end{split}
\end{equation}
The matrices $\mathbf H_{i,j}$ and $\mathbf S_{i,j}$
contain the Hamiltonian and overlap
elements, respectively, between the $i$th and $j$th monomer
units; $a$ is the lattice constant. Furthermore,
we employ a useful algorithm based on transforming Eq.\ref{7}
to a generalised eigenvalue problem for $\lambda\equiv e^{ika}$, yielding
\begin{equation}
\begin{split}
& \left(\begin{array}{cc}
\mathbf H_{m,m}-E\mathbf S_{m,m} & \mathbf H_{m-1,m}-E \mathbf S_{m-1,m} \\
\mathbf I & \mathbf 0 \end{array} \right)
\tilde{\phi}_\lambda \\
& \quad = \lambda \left(\begin{array}{cc}
-(\mathbf H_{m,m+1}-E \mathbf S_{m,m+1}) & \mathbf 0 \\
\mathbf 0 & \mathbf I \end{array} \right)\tilde{\phi}_\lambda.
\label{8}
\end{split}
\end{equation}
The procedure resembles finding the eigenvalues of the transfer matrix
in a non-orthogonal localised basis set but it overcomes
problems arising from possible singular matrices. 
This formal similarity also justifies a relation
of $\beta$ to the complex wavevector. We solve numerically Eq.\ref{8}
for any real $E$ and classify its eigenvalues. For
$\left |\lambda \right|=1$, $k$ is real; otherwise $k$ is complex and
describes exponentially decaying or growing solutions, which are of
particular interest in the tunnelling regime.

\section{Prototype system}
\label{s4}
The molecular junctions under investigation consist of
oligo-phenyl-ethynylene (OPE) wires of small varying length
connected to two graphene electrodes in a ribbon shape.
Before calculating the conductance of systems as that 
depicted in Fig.\ref{f1}, all the formed molecular junctions
are optimised. More precisely, the non-shaded area
is dispatched and the consequent edges are replaced by single hydrogens.
To simulate the bulk electrodes, a constrained atomic
relaxation is performed. Graphene atoms away from the OPE/ribbon
interfacial bond by approximately $3 \times$hexagonal lattice constant
remain immovable. The shaded-area represents infinite graphitic ribbons
whose structure is assumed not to differ from that of
perfect graphene with H saturated edges. As an input
for C-H and single/triple C-C bonds in the graphitic part and
OPE, we use standard literature values. For
the interfacial contact, we take a single C-C bond with $1.39${\AA}.

The choice of ribbons (width as shown in Fig.\ref{f1})
of single-layered graphite is guided by recent
experimental reports \cite{JPC02AM}. In those,
self-assembled-monolayers of conjugated molecules were deposited
on a graphitic substrate. The molecular phenyl rings form a covalent
bond to the carbon electrode in a similar fashion to that of Fig.\ref{f1},
which ensures a low energy barrier at the interface.
Our junctions is a first effort to model nanographite materials
and may be considered as an idealisation with the following approximations:
a) the graphite material is well-ordered, b) the weak
inter-plane interactions in graphite are insignificant,
and c) there is weak inter-molecular interaction. These
assumptions must be studied in detail separately,
which is beyond our present scope.
Later we discuss the validity of our results when other metals are
considered as electrodes.
Our motivation also stems from abiding interest in
carbon-based nanojunctions \cite{PRB01FCR,EPL03GFR}.

\section{Results - Discussion}
\label{s5}
\begin{figure}[h]
\centerline{\epsfig{figure=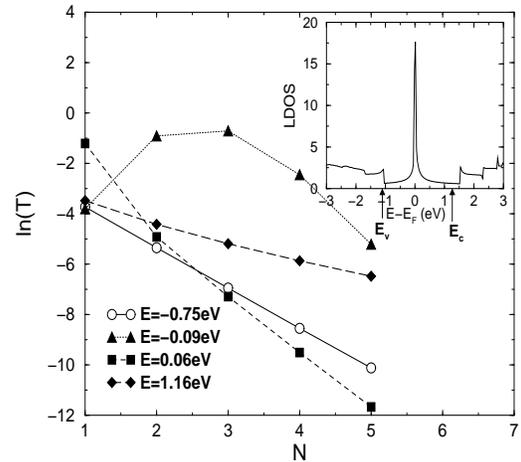,width=15pc,height=15pc}}
\vspace{-0.75cm}
\caption{Logarithm of the transmission function
$T(E)$ as function of the OPE length $N\alpha$, for various
energies $E$. All energies are measured from $E_F$ and lie
in the band-gap of PPE. The inset shows the local DOS
at the contact carbon atom of the graphitic electrode to the oligomer.}
\label{f2}
\end{figure}

We have calculated the conductance spectra
for OPE wires consisting of up to five monomer units
in a large energy window spanning the valence
and conduction bands of poly-phenyl-ethynylene (PPE),
using the formulation of Sec.\ref{s2}. In Fig.\ref{f2},
we plot for selected energies the logarithm of
the transmission function for OPE molecular junctions of increasing
length, represented by the number of monomer
units $N$. The inset shows the local density
of states at the carbon atom of the graphitic electrode
which bonds to the oligomer.

\vspace{-0.5cm}
\vspace{-0.5cm}

There are two types of curves evident in Fig.\ref{f2}.
One type, e.g., at $E=-0.75$eV and $1.16$eV, clearly
indicates an exponential decrease of $T(E)$ with length,
from practically $N=1$. This behaviour
is robust for almost all energies inside the gap formed by
the renormalised HOMO-LUMO energies of the embedded OPE.
This gap is slightly larger than the band-gap of PPE,
allowing for the tunnelling behaviour.
The other type of curve
is approximated by Eq.\ref{2} after a minimum length; for
$E=0.06$eV at $N\approx2$ and for $E=-0.09$eV at $N\approx3$. The latter
data show an anomalous length dependence
with an initial increase of the conductance for $N<3$.
Such behaviour relates to unconventional MIGS \cite{EPL03GFR}.
These provide a way of molecular doping in short junctions
that opens an additional resonant-tunnelling conduction channel.

Unconventional MIGS are manifested when there is a large local DOS
on the metal side. Indeed, the local DOS (inset of Fig.\ref{f2}) at the
electrode interfacial contact of our molecular junction exhibits
(van Hove) singularities. A most pronounced peak lies
within the gap of PPE at the Fermi level of the system. This
corresponds to a state located predominantly at the zigzag edges of the
graphene ribbon \cite{PRB96NFD} and is responsible for the main
departures of $T(E)$ from Eq.\ref{2}. Nevertheless, the tunnel effect
is still observed for quite small chain lengths and energies not in
the immediate vicinity of local DOS peaks ($\sim10$meV), enabling
the extraction of $\beta$. This is a striking result
since it extends the applicability of Eq.\ref{2}
to ultra-short oligomer bridges and to a wide
energy spectrum not far-off to that responsible for resonant-tunnelling.

\begin{figure}[h]
\centerline{\epsfig{figure=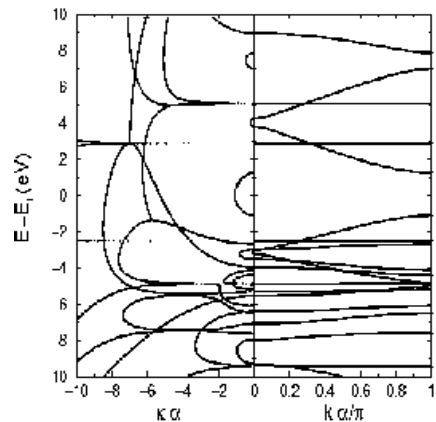,width=14pc,height=14pc}}
\vspace{-0.75cm}
\caption{Complex- and real- band structure of
PPE in left and right panel, respectively.
}
\label{f3}
\end{figure}

We now turn our attention to the electronic structure of perfect PPE.
Its complete band
structure is calculated as described in Sec.\ref{s3}
via Eq.\ref{7} and \ref{8}, and is shown in Fig.\ref{f3}.
The right panel corresponds to the conventional band structure.
The spectrum of the imaginary part of complex $k$
solutions, which come in conjugate pairs ($\pm\kappa$),
is plotted in the left panel. The complex-band structure shows
the typical excursions which connect real bands between local extrema.

An accurate complex-band structure calculation
is necessary for a comprehensive quantitative account of
the MIGS \cite{JCM03PSD}. We have made additional calculations
for alkane chains, graphene, and poly-para-phenylene (PPP). In all
cases, our calculated band structures both in the real
axis and in the complex plane, are in good agreement
with those reported for either minimal
or plane wave basis sets within DFT in the LDA
\cite{PRB02TS,PSSB02TS,PRB00KMS}.
Moreover, as described in Sec.\ref{s3}, MIGS are associated
with the complex $k$ vectors for real energies and
are exponentially decaying. Hence, we anticipate that states
with the smallest $\left | \kappa  \right|$ are dominant. Namely,
of most importance for quantifying the damping factor in Eq.\ref{2}
is the, so-called, real-line connecting the top of the valence and
bottom of the conduction bands, which occur at the border of the
Brillouin zone $k=\pi/\alpha$. This line joins the energies $E_v=-1.12$eV and
$E_c=1.25$eV, and is reproduced quite accurately within DFT-TB.

\vspace{-0.5cm}
\vspace{-0.5cm}

\begin{figure}[h]
\centerline{\epsfig{figure=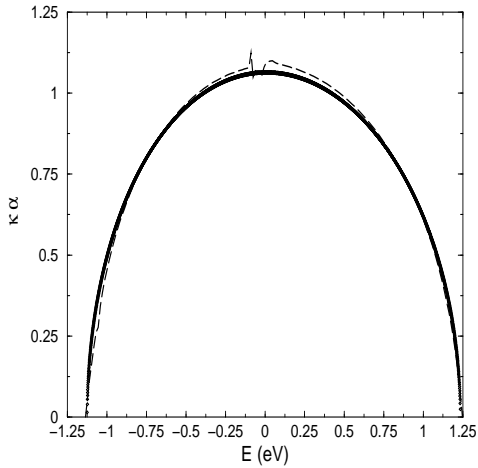,width=15pc,height=15pc}}
\vspace{-0.75cm}
\caption{The $E_c$ to $E_v$ real-line from Fig.\ref{3} (solid)
and the half of the damping factor as extracted from the
slope of transmission function lines (see Fig.\ref{2}) for all
energies (dashed).}
\label{f4}
\end{figure}

In Fig.\ref{f4}, we isolate the positive branch of the
$E_c$ to $E_v$ real-line and plot it on a larger scale.
Together we plot the half of the damping factor $\beta$, which is
numerically obtained by a least-squares fit to determine the
slope of the lines in Fig.\ref{f2}. For energies where $T(E)$
exhibits a non-monotonic length dependence, we have used only
data for $N>2$. The agreement is evidently remarkable for the
entire energy spectrum, even at very small values of the tunnel
decay parameter.

There are hardly deviations from the relation
$\beta(E) = 2 \left | \kappa(E) \right|\alpha $. $\kappa$ denotes
that value of real-lines at $E$ with the largest imaginary part
in the renormalised HOMO-LUMO gap of the oligomer.
Small irregularities appear only at those energies that the
graphitic ribbons have a large local DOS. This marks the specific system
and may not appear if strong disorder in graphene
washes out the singularities.
Since our results point to the intrinsic character of the damping
factor for almost any $\beta(E)N$, we expect an even better agreement
in the usual case of Al or Au electrodes. On the one hand,
such broad-band metals usually exhibit seamless local DOS.
On the other, larger interfacial barriers are formed in
comparison to that of the C-C bond, due to the groups
anchoring the metal. Small departures similar to
those caused by the unconventional MIGS may occur
if the anchor groups have localised states within the spanned
energy window \cite{PRB02TS,PSSB02TS}.

\vspace{-0.5cm}
\vspace{-0.5cm}

Our results generally apply
to oligomer bridges between metals in the off-resonant regime.
They explain qualitatively the microscopic transport mechanism
in experiments of ultra-short oligomer molecular junctions
\cite{APL01SHI,JPCB02WHR,JPCB02IMA},
and results obtained with non-equilibrium Green functions
\cite{PRB03KLG}. For tunnelling, an analysis
\cite{CP01MR} similar to the Simmons model \cite{JAP63S},
but including the spatial discreteness at these length-scales,
may be used to derive current-voltage characteristics. $E_F$ is estimated
by comparison of the extracted damping factor to the
complex wavevector \cite{JPCB02CPZ}.

Conversely, $E_F$ is roughly located at the maximum of the
$E_c$ to $E_v$ real-line for metal-insulator (metal-semiconductor) junctions
\cite{PRB02TS},
which approximately holds for the nanojunctions we studied.
Using this argument, we have calculated the damping factors
for other polymers from the complex wavevector at the middle of
their band-gap. For PPP we found $\beta(E=0)=1.8$,
which is in quantitative agreement with the reported
experimental \cite{JPCB02WHR,JPCB02IMA}
and theoretical \cite{PRB03KLG} values in MMMs of
oligo-para-phenyl chains (OPP) up to four units between Au electrodes.
However, one should take into account the possibility of
non-planar intramolecular phenyl-ring alignment in OPP. We shall
report on such effects in forthcoming studies \cite{FK04} and compare with
other calculations of the complex-band structure for twisted phenyl-ring
geometries \cite{PSSB02TS}.

\section{Concluding remarks}
We demonstrated that no tunnel-barrier collapse
occurs even for {\it ultra-short} oligomer molecular junctions,
and for virtually {\it all energies} within the gap of the corresponding
polymer, $E_c-E_v$. Contrary to expectations, there exists a unique
damping factor $\beta(E)$, even though $\beta(E)N\gg1$ is not always valid.
This is traced back to the larger value, in comparison to the
band-gap $E_c-E_v$, of the renormalised HOMO-LUMO gap of
the oligomers in the junction.

Moreover, we indicated the importance
of calculating, and the need to understand and find ways to engineer,
the complex-band structure of molecular wires. Our results establish
that the damping factor $\beta(E)$ can be almost exactly mapped to
the complex wavevector for all energies within $E_c-E_v$.
This substantiates that the complex-band structure provides a prolific
method to determine the most crucial property of off-resonant electron
transport across oligomer molecular junctions without performing a
transport calculation. Improved independent first-principle methods may
be used to increase the accuracy in evaluating $\beta$.

Our calculations on phenyl-ethynyl chains show that a single complex
wavevector is adequate to explain the tunnelling regime but it cannot be
excluded that a superposition of contributions from few close-by complex-$k$
may be needed to understand electron transport across other oligomers.
Although there are strong indications that our results apply in
many cases, further investigations are clearly necessary to identify the
typical behaviour. In favour of the former are independent calculations on
complex-band structures of various molecular wires, the rapid
convergence of tunnelling decay constants of short oligomers to the long
wire limit (see also \cite{JCP61McC,JPPA94RH}),
and simple qualitative arguments on the sensitivity of transport observables
to small deviations in $\left| \kappa \right|$ due to tunnelling.

The derived connection to the metal-induced gap states fortifies the analysis
of the conductance properties of molecular
junctions by using concepts developed in the theory of
metal-insulator (metal-semiconductor) junctions.
On the other hand, the tunnel properties of oligomers are of special
relevance to molecular electronic devices and
new circuit-designs. Finally, we note the conceptual similarity between
electron transport in metal-molecule-metal junctions
and the fundamental chemical process of molecular electron transfer.
Indeed, a formal analogy in a simple model relates the non-adiabatic
electron transfer rate $k_{DA}$ in a donor-bridge-acceptor system to
the Landauer conductance \cite{Sci03NR}. The expression Eq.\ref{2}
is reminiscent of the superexchange mechanism,
and $\beta$ coincides with McConnell's result \cite{JCP61McC,JPPA94RH}.

\section*{Acknowledgements}
AK is thankful to the
Alexander von Humboldt Stiftung. GF was supported by the
DFG Graduiertenkolleg Nichtlinearit{\"a}t und Nichtgleichgewicht
in kondensierter Materie.
The authors thank Klaus Richter for his critical comments and Rafael
Gutierrez for discussing the manuscript.

\end{document}